\documentclass[a4paper,12pt]{article}

\usepackage{graphics,psfrag}
\usepackage{amsmath,amsthm,amssymb,epsfig,euscript,array,cite,cancel,color}
\usepackage{cases,empheq}
\usepackage[colorlinks=true]{hyperref}     
\usepackage{verbatim}
\usepackage{mciteplus}

\usepackage[usenames,dvipsnames,svgnames,table]{xcolor}

\if{}
\usepackage[paper=a4paper,dvips,top=2.54cm,left=2.74cm,right=2.34cm,
    foot=1cm,bottom=2.54cm]{geometry}
\fi

\theoremstyle{definition}

\theoremstyle{remark}

\newcounter{multieqs}

\newcommand{\be}{\begin{equation}}
\newcommand{\ee}{\end{equation}}
\newcommand{\eq}[1]{(\ref{#1})}
\newcommand{\bit}{\begin{itemize}}  \newcommand{\eit}{\end{itemize}}

\newcommand{\bm}[1]{\mbox{\boldmath $#1$}}
\newcommand{\rf}[1]{(\ref{#1})}

\def\bd{\begin{document}}
\def\ed{\end{document}}
\def\nn{\nonumber}
\def\bea{\begin{eqnarray}}
\def\eea{\end{eqnarray}}
\let\bm=\bibitem

\def\la{\langle}
\def\ra{\rangle}

\def\npb#1#2#3{Nucl. Phys. {\bf{B#1}} #3 (#2)}
\def\plb#1#2#3{Phys. Lett. {\bf{#1B}} #3 (#2)}
\def\prl#1#2#3{Phys. Rev. Lett. {\bf{#1}} #3 (#2)}
\def\prd#1#2#3{Phys. Rev. {D \bf{#1}} #3 (#2)}
\def\cmp#1#2#3{Comm. Math. Phys. {\bf{#1}} #3 (#2)}
\def\cqg#1#2#3{Class. Quantum Grav. {\bf{#1}} #3 (#2)}
\def\nppsa#1#2#3{Nucl. Phys. B (Proc. Suppl.) {\bf{#1A}}#3 (#2)}
\def\ap#1#2#3{Ann. of Phys. {\bf{#1}} #3 (#2)}
\def\ijmp#1#2#3{Int. J. Mod. Phys. {\bf{A#1}} #3 (#2)}
\def\rmp#1#2#3{Rev. Mod. Phys. {\bf{#1}} #3 (#2)}
\def\mpla#1#2#3{Mod. Phys. Lett. {\bf A#1} #3 (#2)}
\def\jhep#1#2#3{J. High Energy Phys. {\bf #1} #3 (#2)}
\def\atmp#1#2#3{Adv. Theor. Math. Phys. {\bf #1} #3 (#2)}

\def\N{{\cal N}}
\def\sst{\scriptscriptstyle}
\def\thetabar{\bar\theta}
\def\Tr{{\rm Tr}}
\def\one{\mbox{1 \kern-.59em {\rm l}}}

\def\a{\alpha}      \def\da{{\dot\alpha}}  \def\dA{{\dot A}}
\def\b{\beta}       \def\db{{\dot\beta}}
\def\g{\gamma}  \def\G{\Gamma}  \def\dc{{\dot\gamma}}
\def\d{\delta}  \def\D{\Delta}  \def\ddt{\dot\delta}
\def\e{\epsilon}        \def\ve{\varepsilon}
\def\f{\phi}    \def\F{\Phi}    \def\vvf{\f}
\def\h{\eta}
\def\k{\kappa}
\def\l{\lambda} \def\L{\Lambda}
\def\m{\mu} \def\n{\nu}
\def\o{\omega}
\def\p{\pi} \def\P{\Pi}
\def\r{\rho}
\def\s{\sigma}  \def\S{\Sigma}
\def\t{\tau}
\def\th{\theta} \def\Th{\Theta} \def\vth{\vartheta}
\def\X{\Xeta}
\def\z{\zeta}

\def\na{\nabla}

\def\cA{{\cal A}} \def\cB{{\cal B}} \def\cC{{\cal C}}
\def\cD{{\cal D}} \def\cE{{\cal E}} \def\cF{{\cal F}}
\def\cG{{\cal G}} \def\cH{{\cal H}} \def\cI{{\cal I}}
\def\cJ{{\cal J}} \def\cK{{\cal K}} \def\cL{{\cal L}}
\def\cM{{\cal M}} \def\cN{{\cal N}} \def\cO{{\cal O}}
\def\cP{{\cal P}} \def\cQ{{\cal Q}} \def\cR{{\cal R}}
\def\cS{{\cal S}} \def\cT{{\cal T}} \def\cU{{\cal U}}
\def\cV{{\cal V}} \def\cW{{\cal W}} \def\cX{{\cal X}}
\def\cY{{\cal Y}} \def\cZ{{\cal Z}}

\def\ua{\underline{\alpha}}
\def\uc{\underline{\phantom{\alpha}}\!\!\!\gamma}
\def\um{\underline{\mu}}
\def\ud{\underline\delta}
\def\ue{\underline\epsilon}
\def\una{{\underline a}}\def\unA{\underline A}
\def\unb{{\underline b}}\def\unB{\underline B}
\def\unc{{\underline c}}\def\unC{\underline C}
\def\und{{\underline d}}\def\unD{\underline D}
\def\une{{\underline e}}\def\unE{\underline E}
\def\unf{\underline{\phantom{e}}\!\!\!\! f}\def\unF{\underline F}
\def\unm{{\underline m}}\def\unM{\underline M}
\def\unn{{\underline n}}\def\unN{\underline N}
\def\unp{\underline{\phantom{a}}\!\!\! p}\def\unP{\underline P}
\def\unq{\underline{\phantom{a}}\!\!\! q}
\def\unQ{\underline{\phantom{A}}\!\!\!\! Q}
\def\unH{{\underline{H}}}
\def\unl{{\underline{l}}}

\def\As {{A \hspace{-6.4pt} \slash}\;}
\def\bs {{b \hspace{-6.4pt} \slash}\;}
\def\Ds {{D \hspace{-6.4pt} \slash}\;}
\def\Gts {{\Gt \hspace{-6.4pt} \slash}\;}
\def\ds {{\del \hspace{-6.4pt} \slash}\;}
\def\ss {{\s \hspace{-6.4pt} \slash}\;}
\def\ks {{ k \hspace{-6.4pt} \slash}\;}
\def\ps {{p \hspace{-6.4pt} \slash}\;}
\def\xs {{x \hspace{-6.4pt} \slash}\;}
\def\pas {{{p_1} \hspace{-6.4pt} \slash}\;}
\def\pbs {{{p_2} \hspace{-6.4pt} \slash}\;}
\def\cFs {{{\cal F} \hspace{-6.4pt} \slash}\;}

\def\Ah{{\hat{A}}}
\def\Dh{{\hat{D}}}
\def\Gh{{\hat{G}}}
\def\Fh{{\hat{F}}}
\def\Ih{{\hat{I}}}
\def\Jh{{\hat{J}}}
\def\Kh{{\hat{K}}}
\def\Lh{{\hat{L}}}
\def\Ph{{\hat{P}}}
\def\Rh{{\hat{R}}}
\def\Vh{{\hat{V}}}
\def\Xh{{\hat{X}}}

\def\ah{{\hat{a}}}
\def\bh{{\hat{b}}}
\def\ch{{\hat{c}}}
\def\gh{{\hat{g}}}
\def\dh{{\hat{d}}}
\def\hh{{\hat{h}}}
\def\uh{{\hat{u}}}
\def\vh{{\hat{v}}}
\def\xh{{\hat{x}}}
\def\yh{{\hat{y}}}
\def\zh{{\hat{z}}}
\def\ph{{\hat{p}}}
\def\thh{{\hat{t}}}
\def\xih{\hat{\xi}}
\def\Psih{\hat{\Psi}}
\def\mh{{\hat{m}}}
\def\nh{{\hat{n}}}
\def\ih{{\hat{i}}}
\def\jh{{\hat{j}}}
\def\kh{{\hat{k}}}
\def\aah{{\hat{\alpha}}}
\def\bbh{{\hat{\beta}}}
\def\ggh{{\hat{\gamma}}}
\def\llh{{\hat{\ell}}}
\def\ph{{\hat{p}}}

\def\psit{\tilde{\psi}}
\def\Psit{\tilde{\Psi}}
\def\Psibt{\tilde{\bar{Psi}}}

\def\delt{\tilde{\delta}}
\def\Phit{\tilde{\Phi}}
\def\Phitb{\overline{\tilde{Phi}}}
\def\tht{\tilde{\th}}
\def\lt{\tilde{\l}}
\def\chit{\tilde{\chi}}
\def\phit{\tilde{\phi}}

\def\At{\tilde{A}}
\def\Bt{\tilde{B}}
\def\Ct{\tilde{C}}
\def\Dt{\tilde{D}}
\def\Et{\tilde{E}}
\def\Ft{\tilde{F}}
\def\Gt{\tilde{G}}
\def\Ht{\tilde{H}}
\def\It{\tilde{I}}
\def\Jt{\tilde{J}}
\def\Qt{\tilde{Q}}
\def\Rt{\tilde{R}}
\def\Mt{\tilde{M }}
\def\Nt{\tilde{N}}
\def\St{\tilde{S}}
\def\Vt{\tilde{V}}
\def\Xt{\tilde{X}}
\def\at{\tilde{a}}
\def\ct{\tilde{c}}
\def\dt{\tilde{d}}
\def\htt{\tilde{h}}
\def\ft{\tilde{f}}
\def\gt{\tilde{g}}
\def\pt{\tilde{p}}
\def\qt{\tilde{q}}
\def\vt{\tilde{v}}
\def\nt{\tilde{n}}
\def\ut{\tilde{u}}
\def\wt{\tilde{w}}
\def\zt{\tilde{z}}
\def\xt{\tilde{x}}
\def\yt{\tilde{y}}
\def\Psit{\tilde{\Psi}}
\def\vphit{\tilde{\varphi}}
\def\Tt{{\tilde{T}}}

\def\eb{\bar{\epsilon}}
\def\delb{\bar{\partial}}
\def\thb{\bar{\theta}}
\def\mub{\bar{\mu}}
\def\lamb{\bar{\l}}
\def\psib{\bar{\psi}}
\def\sb{\bar{\sigma}}
\def\xib{\bar{\xi}}
\def\chib{\bar{\chi}}

\def\Psib{\bar{\Psi}}
\def\Phib{\bar{\Phi}}
\def\Lamb{\bar{\Lambda}}
\def\Sb{{\overline \Sigma}}
\def\cb{\bar{c}}
\def\hb{\bar{h}}
\def\qb{\bar{q}}
\def\wb{\bar{w}}
\def\ub{\bar{u}}
\def\zb{{\bar{z}}}
\def\Hb{\bar{H}}
\def\Qb{{\bar Q}}
\def\Omegab{\overline{\Omega}}
\def\ob{\overline{\omega}}

\def\Ab{{\overline A}} \def\Bb{{\overline B}} \def\Cb{{\overline C}}
\def\Db{{\overline D}} \def\Eb{{\overline E}} \def\Fb{{\overline F}}
\def\Gb{{\overline G}}
\def\Ib{{\overline I}}
\def\Jb{{\overline J}} \def\Kb{{\overline K}} \def\Lb{{\overline L}}
\def\Mb{{\overline M}} \def\Nb{{\overline N}} \def\Ob{{\overline O}}
\def\Pb{{\overline P}}  \def\Rb{{\overline R}}
 \def\Tb{{\overline T}} \def\Ub{{\overline U}}
\def\Vb{{\overline V}} \def\Wb{{\overline W}} \def\Xb{{\overline X}}
\def\Yb{{\overline Y}} \def\Zb{{\overline Z}}

\def\fb{{\overline f}}
\def\gb{{\overline g}}
\def\mb{{\overline m}}
\def\lb{{\overline l}}
\def\yb{{\overline y}}

\def\ldel{{\overleftarrow{\del}}}
\def\rdel{{\overrightarrow{\del}}}
\def\ldeldel{{\overleftarrow{\del^2}}}
\def\rdeldel{{\overrightarrow{\del^2}}}
\def\ldelb{{\overleftarrow{\bar{\del}}}}
\def\rdelb{{\overrightarrow{\bar{\del}}}}

\def\ba{{\bf a}}
\def\bk{{\bf k}}
\def\bl{{\bf l}}
\def\bp{{\bf p}}
\def\bq{{\bf q}}
\def\br{{\bf r}}
\def\bt{{\bf t}}
\def\bu{{\bf u}}
\def\bv{{\bf v}}
\def\bx{{\bf x}}
\def\by{{\bf y}}
\def\bR{{\bf R}}
\def\bV{{\bf V}}

\def\bone{{\bf 1}}

\def\va{{\vec a}}
\def\vk{{\vec k}}
\def\vp{{\vec p}}
\def\vq{{\vec q}}
\def\vx{{\vec x}}
\def\vy{{\vec y}}
\def\vu{{\vec u}}
\def\vv{{\vec v}}

\def\vs{{\vec \sigma}}
\def\vtau{{\vec \tau}}

\newcommand{\ov}[1]{\overrightarrow{#1}}

\def\frA{\mathfrak{A}}
\def\frB{\mathfrak{B}}
\def\frC{\mathfrak{C}}
\def\frD{\mathfrak{D}}
\def\frE{\mathfrak{E}}
\def\frF{\mathfrak{F}}
\def\frG{\mathfrak{G}}
\def\frH{\mathfrak{H}}
\def\frM{\mathfrak{M}}
\def\frN{\mathfrak{N}}
\def\frR{\mathfrak{R}}
\def\frW{\mathfrak{W}}

\def\fra{\mathfrak{a}}
\def\frb{\mathfrak{b}}
\def\frf{\mathfrak{f}}
\def\frg{\mathfrak{g}}
\def\frh{\mathfrak{h}}
\def\frl{\mathfrak{l}}
\def\frs{\mathfrak{s}}
\def\fri{\mathfrak{i}}
\def\frj{\mathfrak{j}}

\def\ma{\mathfrak{a}}
\def\mg{\mathfrak{g}}
\def\mh{\mathfrak{h}}
\def\mR{\mathfrak{R}}
\def\mN{\mathfrak{N}}

\def\d{\delta}\def\D{\Delta}\def\ddt{\dot\delta}

\def\pa{\partial} \def\del{\partial}
\def\xx{\times}
\def\uno{\mbox{1 \kern-.59em {\rm l}}}

\def\trp{^{\top}}
\def\inv{^{-1}}
\def\dag{{^{\dagger}}}
\def\pr{^{\prime}}

\def\rar{\rightarrow}
\def\lar{\leftarrow}
\def\lrar{\leftrightarrow}

\newcommand{\0}{\,\!}      
\def\one{1\!\!1\,\,}
\def\im{\imath}
\def\jm{\jmath}

\newcommand{\tr}{\mbox{tr}}
\newcommand{\slsh}[1]{/ \!\!\!\! #1}

\def\vac{|0\rangle}
\def\lvac{\langle 0|}

\def\hlf{\frac{1}{2}}
\def\ove#1{\frac{1}{#1}}

\def\Box{\square}
\def\CC {\mathbb{C}}
\def\FF {\mathbb{F}}
\def\RR{\mathbb{R}}
\def\NN{\mathbb{N}}
\def\ZZ{\mathbb{Z}}
\def\bb#1{{\bf #1}}
\def\bcomment#1{}
\def\bfhat#1{{\bf \hat{#1}}}
\def\VEV#1{\left\langle #1\right\rangle}

\newcommand{\ex}[1]{{\rm e}^{#1}} \def\ii{{\rm i}}

\newcommand{\lrbrk}[1]{\left(#1\right)}
\newcommand{\lrsbrk}[1]{\left[#1\right]}
\newcommand{\sfrac}[2]{{\textstyle\frac{#1}{#2}}}

\def\stw{{\sqrt{2}}}

\def\rf {{\rm f}}
\def\ri {{\rm i}}
\def\rj {{\rm j}}
\def\rk {{\rm k}}
\def\rl {{\rm l}}
\def\rs {{\scriptscriptstyle \rm S}}
\def\rt {{\scriptscriptstyle \rm T}}

\def\rQ {{\scriptscriptstyle \rm \cQ}}
\def\rR {{\scriptscriptstyle \rm \cR}}

\def\cQb{{\cal \Qb}}
\def\cRb{{\cal \Rb}}
\def\cWb{{\cal \Wb}}

\def\fd {{\rm N}}
\def\afd {{\overline{\rm N}}}

\def \II {I\hspace{-.1em}I\hspace{.1em}}
\def \IIA {\mbox{\II A\hspace{.2em}}}
\def \IIB {\mbox{\II B\hspace{.2em}}}
\def \gs {g^s}
\def \ls {\lambda^s}

\def \I {{\cal I}}
\def \qs {q\hspace{-.53em}/\hspace{.15em}}
\def \ks {k\hspace{-.53em}/\hspace{.15em}}
\def \YM {{\mbox{\tiny YM}}}
\def \gym {g_{\YM}}

\def \Lc {\L_c}
\def\IR{\relax{\rm I\kern-.18em R}}
\def \id {{\bf 1}}

\def\cci{\ell}
\def\ccj{\ell'}

\def \thbb{\overline{\th\th}}
\newcommand \ol{\overline}
\def \lamb{\bar{\lambda}}
\def \vphi{\varphi}
\def \lambh{\hat{\bar{\lambda}}}
\def \lh{\hat{\lambda}}
\def \dd{\ddagger}

\def \ad {\dot{a}}
\def \bd {\dot{b}}
\def \cd {\dot{c}}
\def  \ddd {\dot{d}}
\def \ed {\dot{e}}
\def \fd {\dot{f}}
\def \Bh {\hat{B}}
\def \zm {{(0)}}
\def \nz {{(\text{KK})}}
\def \3{{(3)}}
\def \diag {\text{diag}}
\def \inm {{(m^{-1})}}

\def\eh{{\hat{e}}}
\def\fh{{\hat{f}}}
\renewcommand{\mh}{\hat{m}}
\def\theequation{\thesection.\arabic{equation}}

\def\adj{\text{adj}}
\def\co{\text{co}}
\def\6{{\text{(6)}}}
\def\5{{\text{(5)}}}
\def\Hbt{\tilde{\bar{H}}}

\author{Sheng-Lan Ko\footnote{sheng-lank@nu.ac.th}~
and  Pichet Vanichchapongjaroen\footnote{pichetv@nu.ac.th}$~$
\\
\\
{\small  
	\it The Institute for Fundamental Study ``The Tah Poe Academia Institute",}
\\
{\small\it Naresuan University, Phitsanulok 65000, Thailand}
}

\title{\bf The Dual Formulation of M5-brane Action}

\date{}
\begin{document}
\maketitle

\abstract{ 
We construct a dual formulation, with respect to the conventional PST formalism, 
of the M5--brane action propagating in a generic 11d supergravity background. 
Constraint analysis is performed to further justify that our theory has 
the correct number of degrees of freedom. 
Comparison of this action with the existing M5--brane actions is carried out. 
We also show that a conventional D4--brane action is obtained upon 
double dimensional reduction. 
}

\thispagestyle{empty}
\newpage
\tableofcontents

\setcounter{equation}0
\section{Introduction}

    In the literature, two main approaches are used to
    describe dynamics of M5--brane theory.
    The first approach is called the superembedding approach \cite{Sorokin:1999jx},
    in which a supersymmetric M5--brane worldvolume is embedded into
    an 11--dimensional supersymmetric target space. This approach was carried out in \cite{Howe:1996yn,Howe:1997fb}. 
    Alternatively, the approach called the Green--Schwarz approach,
    in which a bosonic M5--brane worldvolume is embedded into
    an 11--dimensional supersymmetric target space can also be used.
    The first successful attempt on this approach is shown in \cite{Pasti:1997gx,Bandos:1997ui,Aganagic:1997zq},
    in which the action known as the PST action is constructed.
    Within the Green--Schwarz approach, it is also possible to construct alternative actions which are expected to serve some specific purposes better. 
    In particular, with the hope to understand the connection between
    the five-brane proposal \cite{Ho:2008nn,Ho:2008ve} and 
    the known complete M5 action,
    the alternative action \cite{Pasti:2009xc, Ko:2013dka} is constructed. Although it is yet unclear whether
    this alternative action would eventually serve its original intended purpose,
    the possibility to have more than one action which fully describes
    a single supersymmetric M5--brane should already be a good motivation
    to seek further alternative actions. 
    In this paper, we construct yet another alternative action.
    An attempt toward the 2+4 formulation of M5-brane action 
    is put forward in \cite{Ko:2015zsy}. However, it is still unclear 
    whether the completion of \cite{Ko:2015zsy} to an M5-brane is possible. 

Making symmetries of underlying theory manifest is always beneficial, and 
it is no exception for duality symmetry. 
For example, the construction of the duality symmetric action of 11d supergravity 
allows its direct coupling to 
both the M2-- and M5--branes \cite{Bandos:1997gd}. 
For us, we are interested in the duality--symmetric worldvolume action for the M5--brane. 
In \cite{Maznytsia:1998xw,Maznytsia:1998yr}, in order to reflect the duality property of M5--brane, 
  the dual action to the quadratic PST action \cite{Pasti:1996vs} for chiral 2-form
  is constructed. This action, however, still does not describe the M5--brane
  as the non-linearisation and the couple to the other fields have yet to be implemented.
  The main goal in our paper is to obtain the M5--brane action
  in the dual formalism.

  In order to achieve this goal,
  we start from considering the quadratic action for the chiral 2-form in the dual formulation.
  In a sense, this is obtained from using the gauge freedom to fix the auxiliary field
  of the model in \cite{Maznytsia:1998xw,Maznytsia:1998yr}. We have checked that despite the fact that
  the Lorentz symmetry is not manifest due to a certain space--like direction is singled out,
  this action can be shown to have the modified version of Lorentz symmetry.
  We next couple this action with 6d gravity and show that it 
  possesses
  a modified version of diffeomorphism symmetry.
  Next, we non-linearise the action by utilising
  the idea of \cite{DeCastro:2002vd, DeCastro:2001he,DeCastro:1998pm},
  in which one starts from the known Hamiltonian for the gauge-fixed 
  PST action,
  relax certain constraints, and then work out the action.
  Having obtained the linearised action, the extension to the M5--brane action
  is straightforward. The most non-trivial check is whether the action
  possesses the kappa symmetry. We have shown that this is the case.
          
          In \cite{Pasti:1997gx,Aganagic:1997zq}, it was shown that the double dimensional reduction of the gauge-fixed PST M5--brane action gives rise to 
          the dual D4--brane action \cite{Aganagic:1997zk}. With the dual nature of 
          the dual 1+5 M5 action, it is anticipated that a standard D4 action \cite{Douglas:1995bn, Green:1996bh} written in terms of a worldvolume vector gauge field would be 
          obtained upon double dimensional reduction. 
          We have carried out the dimensional reduction and found that it is indeed 
          the case.

   The paper is organized as follows. In Section \ref{sec:M5Ss},
  we first review the M5--brane action in the PST formulation, 
  and then present the M5--brane action in the dual 1+5 formulation.
  Its derivation is shown in Section \ref{sec:deri}. In Section \ref{sec:constr}, the constraint analysis
  of the action is discussed.
  This presents further verifications on the action.
  In Section \ref{sec:osS}, we show that the on-shell values of the dual 1+5 action
  equals to those of the 1+5 and 3+3 actions.
  It is shown in Section \ref{sec:DDR} that the double dimensional reduction of dual 1+5 M5 action gives directly the standard D4 action. 
  In Conclusion we summarise our results and discuss some 
  open problems and possible future works.

\setcounter{equation}0
\section{The M5-brane actions}  \label{sec:M5Ss}

The action for a single supersymmetric M5-brane
in the Green--Schwarz approach describes
an M5--brane embedded into an 11 dimensional
target superspace.
Within this approach its first formulation, known as the PST formulation,
is presented in \cite{Bandos:1997ui, Aganagic:1997zq}.
In the PST formulation, an auxiliary scalar field is introduced.
There is a local gauge symmetry which reflects the auxiliary nature of this field.
After a gauge fixing of this symmetry, the auxiliary field is
identified as one of the coordinates of the 6d worldvolume.
As a result,
the 6d worldvoulme indices are separated into 1d and 5d ones.
The 6d covariance is therefore no longer manifest.
However, it can be shown that the resulting theory still has the full 6d diffeomorphism symmetry,
which is modified.
Furthermore,
as in the case of the original PST formulation,
the resulting action legitimately describes a single supersymmetric M5--brane
in a generic 11d supergravity superbackground.
Therefore due to the way the indices are separated,
we call this action, which is a result of the gauge-fixing of the auxiliary field
from the one in PST formulation,
as being in the 1+5 formulation.
Note that in this formulation, the singled out direction
can either be space--like or time--like.
In particular, we will call the 1+5 formulation in which
the space--like direction is singled out as the PS1+5 formulation,
whereas the one in which the time--like direction is singled out
will be called the HT1+5 formulation.

More recently it is shown by construction in \cite{Ko:2013dka}
that, also within the Green--Schwarz approach, there exists
an alternative formulation of a single supersymmetric M5-brane.
In this formulation, there are three auxiliary scalar fields.
After the gauge--fixing of these scalar fields,
the 6d worldvolume indices are separated into 3d and 3d ones.
The resulting theory also legitimately describes a single supersymmetric M5--brane,
and is said as being in the 3+3 formulation.
Having an extra formulation at hand,
it is natural to expect that this
would eventually prove useful in order to understand more about the nature of M5-brane,
and of course it would be natural to seek for other formulations.
In \cite{Ko:2015zsy}, an attempt was made in order to
construct the 2+4 formulation. However, it has not yet been clear
whether such a construction would be possible.

In this paper, we construct and present yet another formulation,
called the dual 1+5 formulation.
To the best of our knowledge, the complete M5--brane action in the
dual 1+5 formulation has not been presented nor discussed
before in the literature.
As for the case of the 1+5 formulation,
one direction on the 6d worldvolume is singled out.
However, the singled out direction
can only be space--like.
This feature is different from the 1+5 formulation,
in which the singled out direction can either be space--like
or time--like.

In this paper, the signature of the metric of the 
11--dimensional target superspace is taken to be mostly plus.
It is parametrized by $Z^{\mathcal M}=(X^M,\th)$, 
in which $X^M$ are eleven bosonic coordinates 
and $\th$ are 32 real fermionic coordinates. 
The geometry of the 11d supergravity are described by tangent-space vector super-vielbeins $E^A(Z)=dZ^{\cM}E_{\cM}{}^A(Z)$ ($A=0,1,2,\cdots, 10)$ and Majorana-spinor super-vielbeins $E^\alpha(Z)=dZ^{\cM}E_{\cM}{}^\alpha(Z)$ ($\a=1,2,\cdots, 32)$.

The vector super-vielbein satisfies the following essential torsion constraint, which is required for proving the kappa-symmetry of the $M5$-brane action,
\be\label{T}
T^A=DE^A=dE^A+E^B\Omega_B{}^A=-iE^\a \Gamma^A_{\a\b}E^\beta\,,
\ee
where $\Omega_B{}^A(Z)$ is the 1-form spin connection in eleven dimension, $\Gamma^A_{\a\b}=\Gamma^A_{\b\a}$ are real symmetric gamma matrices and the exterior differential acts from the right.

The coordinates $x^\mu$ ($\mu=0,1,\cdots, 5$) parametrize the worldvolume of the M5-brane which carries the chiral 2-form gauge field 
$B_2(x)=\ove{2}dx^\mu dx^\nu B_{\n\m}(x)$. 
The induced metric on the $M5$-brane worldvolume is constructed with the pull-backs of the vector super-vielbeins $E^A(Z)$
\be\label{g6}
g_{\mu\nu}(x)=E_\m^AE^B_\n
\eta_{AB},
\qquad E_\mu^A=\partial_\m Z^{\cN}E_{\cN}{}^A(Z(x)).
\ee 

The $M5$-brane couples to the 11d supergravity 3-form gauge superfield, $C_3(Z)=\frac 1{3!}
dZ^{\cM_1}dZ^{\mathcal M_2}dZ^{\mathcal M_3}
C_{\cM_3  \cM_2  \cM_1}$,
and its $C_6(Z)$ dual.
Their field strengths are constrained as follows
\be \label{F47}
\begin{split}
dC_3
&= -\frac i2 E^AE^BE^\alpha E^\beta(\Gamma_{BA})_{\alpha\beta}+\frac 1{4!} E^AE^BE^CE^DF^{(4)}_{DCBA}(Z)\,, \\
dC_6-C_3dC_3
&= \frac{2i}{5!} E^{A_1}\cdots E^{A_5}E^\alpha E^\beta(\Gamma_{A_5\cdots A_1})_{\alpha\beta}+\frac 1{7!} E^{A_1}\cdots E^{A_7}F^{(7)}_{A_7\cdots A_1}(Z)\,\\
F^{(7)\,A_1\cdots A_7}
&= \frac 1{4!}\epsilon^{A_1\cdots A_{11}}F^{(4)}_{A_8\cdots A_{11}}\,, \qquad \epsilon^{0...10}=-\epsilon_{0...10}=1. 
\end{split}
\ee
The extended field strengths of $B_2(x)$ which appears in the M5-brane action is
\be\label{H3}
H_3=dB_2+C_3\,,
\ee
where $C_3(Z(x))$ is the pullback of the 3-form gauge field on the M5-brane worldvolume .

Having discussed the background set-up, 
we next proceed by briefly reviewing the original form of the M5--brane action and then will present our main result, namely, the alternative worldvolume action for the M5--brane in a generic $D=11$ supergravity background.

\subsection{Original M5--brane action}
In this case to ensure the $6d$ worldvolume covariance of the $M5$--brane action one uses an auxiliary scalar field $a(x)$, whose gradient $\pa_\m a$ could be either time--like, i.e. in a certain gauge, $\pa_\m a = \d_\m^0$ or space--like $\pa_\m a = \d_\m^5$.

The $M5$--brane action in a generic $D=11$ supergravity superbackground constructed in \cite{Pasti:1997gx,Bandos:1997ui,Aganagic:1997zq} has the following form:
\bea \label{PSTM5}
S
&=& -\int_{\mathcal{M}_6} d^6x \lrsbrk{\sqrt{ - \det\lrbrk{g_{\m\n} + i\frac{\partial^\r a}{\sqrt{(\del a)^2}} \Hbt_{\r\m\n}}}  + \frac{\sqrt{-g}}{4(\del a)^2} \del_\l a \Hbt^{\l\m\n}H_{\m\n\r}\del^\r a } \nn\\
&&
 + \ove{2}\int_{\mathcal{M}_6} \lrbrk{ C_6 + H_3 \wedge C_3 },
\eea
with
\be   \label{oldHbt}
\Hbt^{\r\m\n} \equiv \frac{1}{6\sqrt{-g}} \, \e^{\r\m\n\l\s\t} H_{\l\s\t},\quad
g=\det g_{\m\n}\,,
\ee
where
$$
\epsilon^{0\cdots 5}=-\epsilon_{0\cdots 5}=1\,.
$$

By utilising local gauge symmetries,
it is possible to set $a = x^5.$
The action then becomes
\bea \label{PSTM5noncov}
S
&=& - \int_{\mathcal{M}_6} d^6x \lrsbrk{\sqrt{ - \det(g_{\m\n} + i(\Hbt\cdot u)_{\m\n})}  + \frac{\sqrt{-g}}{4} (\Hbt\cdot u)^{\m\n}(H\cdot u)_{\m\n} } \nn\\
&&
 + \ove{2}\int_{\mathcal{M}_6} \lrbrk{ C_6 + H_3 \wedge C_3 },
\eea
with
\be
(\Hbt\cdot u)_{\m\n} \equiv \Hbt_{\m\n\r}u^\r,\qquad
(H\cdot u)_{\mu\nu} \equiv H_{\mu\nu\rho} u^\r
\ee
where
\be
u_\l \equiv \frac{\d^5_\l}{\sqrt{g^{55}}},\qquad
u^\l \equiv \frac{g^{\l 5}}{\sqrt{g^{55}}},\qquad
\ee

In the action \eq{PSTM5noncov},
the 6d indices on the M5-brane worldvolume
are separated into the 5d indices and the index $5.$
The 6d indices are represented by
the Greek letters $\m,\n,\cdots = 0,1,2,3,4,5$ while 
the 5d indices are represented by
the underlined latin indices $\una,\unb,\cdots = 0,1,2,3,4$.
Despite the explicit separation of the indices,
the action still possess the diffeomorphism symmetries.
See \cite{Henneaux:1988gg,Schwarz:1997mc} for example.

In addition to the conventional abelian gauge symmetry for the chiral 2-form,
the action (\ref{PSTM5noncov}) has the following local gauge symmetry:
\be \label{15SYM1}
\d B_{\una\unb} = 0,\qquad
\d B_{5\una} = \Phi_{\una}(x),
\ee
with 
$\Phi_\una(x)$ being arbitrary local functions on the woldvolume.
The symmetry (\ref{15SYM1}) ensures that the equation of motion of $B_2$ reduces to the non--linear self--duality condition
\be\label{sdM5o}
(H\cdot u)_{\mu\nu}= \cU_{\mu\nu}(\Hbt)\,,
\ee
where
\be\label{calV}
\cU^{\mu\nu}(\Hbt)
\equiv -2\,\frac{\d \sqrt{\det(\delta^\nu_{\mu} + i(\Hbt\cdot u)_{\mu}{}^{\nu})}}{\d (\Hbt\cdot u)_{\mu\nu}}.
\ee

The action \eqref{PSTM5noncov} is also invariant under the local fermionic kappa--symmetry transformations with the parameter $\kappa^\alpha(x)$ which acts on the pullbacks of the target--space supervielbeins and the $B_2$ field strength as follows
\bea\label{PSTkappa}
i_\k E^\alpha \equiv \delta_\k Z^{\mathcal M} E^\alpha_{\mathcal M} = \frac 12 (1+\bar\Gamma)^\alpha{}_{\beta} \k^{\beta}, \quad
i_\k E^A \equiv \delta_\k Z^{\mathcal M} E^A_{\mathcal M}  =  0. \\
\delta g_{\mu\nu} = -4i E^\alpha_{(\mu} (\G_{\nu)})_{\alpha\beta} \, i_\k E^{\beta}, \quad
\d{H}^\3 = i_\k d C^\3, \quad \delta_\k a(x) = 0\,,    \nn
\eea
where $(1+\bar\Gamma)/2$ is the projector of rank 16 with $\bar \Gamma$ having the following form
\bea\label{barGPST}
\!\!\!\!\!\!\!\!\!\sqrt{\det(\delta_{\mu}^{\nu} + i(\Hbt\cdot u)_{\mu}{}^{\nu})} \,\bar\Gamma
&=& \gamma^{(6)}
- \frac{1}{2}\G^{\mu\nu\lambda} u_{\mu}(\Hbt\cdot u)_{\nu\lambda}
 - \ove{16\sqrt{-g}} {\e^{\mu_1\cdots\mu_6}}
		(\Hbt\cdot u)_{\mu_1\mu_2}(\Hbt\cdot u)_{\mu_3\mu_4} \G_{\mu_5\mu_6}, \nn\\
\bar\Gamma^2&=& 1\,,\qquad \tr{\bar\Gamma}=0,
\eea
where
\be\label{gamma6}
\Gamma_\mu=E_\mu{}^A\Gamma_A\,,\qquad \gamma^{(6)}=\frac 1{6!\sqrt{-g}}\epsilon^{\mu_1\cdots\mu_6}\Gamma_{\mu_1\cdots\mu_6}\,.
\ee

\subsection{M5--brane action in the dual formulation}

In this paper, we construct an M5--brane action
in the dual formulation and show that it has all the required properties, 
that is it is self-interacting, diffeomorphism invariant 
and kappa symmetric.
Let us first present the action along with basic discussions.
\bea\label{dual15action}
S
&=& \int_{\mathcal{M}_6} d^6x \lrsbrk{-\sqrt{-g}\sqrt{\det\lrbrk{\d_\m^\n + (H\cdot v)_\m{}^\n } }  + \frac{\sqrt{-g}}{4} (\Hbt\cdot v)^{\m\n}(H\cdot v)_{\m\n} } \nn\\
&&
 +\hlf \int_{\mathcal{M}_6} \lrbrk{ C_6 + H_3 \wedge C_3 },
\eea
with 
\be
(\Hbt\cdot v)_{\m\n} \equiv \Hbt_{\m\n\r}v^\r,\qquad
(H\cdot v)_{\mu\nu} \equiv H_{\mu\nu\rho} v^\r,
\ee
where
\be
v_\l \equiv \frac{g_{5\l}}{\sqrt{g_{55}}},\qquad
v^\l \equiv \frac{\d^\l_5}{\sqrt{g_{55}}}.
\ee
This theory has the semi-local gauge symmetry
\be
\d B_{\una\unb} = \o_{\una\unb}(x^\unl),\qquad
\d B_{\una 5} = 0,
\ee
where $\omega_{\underline{ij}} = \omega_{\underline{[ij]}}(x^{\underline{l}})$ are arbitrary 
functions of 5d coordinates $x^{\underline{l}}$.
This semi-local gauge symmetry
can be used to ensure that the equation of motion of $B_2$
reduces to the non-linear self-duality condition
\be  \label{nlSD}
-\ove{3} \ove{g_{55}} g_{5[5} 
			\e^{\underline{ablmn}} H_{\underline{lmn}]} 
			+ 3\frac{\d V}{\d H_{\una\unb5}}
= 0, 
\ee
where
\be
V = V(g_{\m\n},H_{5\una\unb})= - \sqrt{-g}\sqrt{\det\lrbrk{\d_{\una}^{\unb} + \frac{1}{\sqrt{g_{55}}}H_{5\una}{}^{\unb}}}
\ee

This alternative M5-brane action is also invariant under the kappa symmetry \eqref{PSTkappa} with 
\be
\sqrt{\det(\d_\m^\n + H_\m{}^\n)}\bar\G
=\g^\6 + \hlf v_\a H_{\b\g}\g^\6\G^{\a\b\g} + \ove{16\sqrt{-g}}\e^{\m_1\cdots\m_6}H_{\m_1\m_2}H_{\m_3\m_4}
\G_{\m_5\m_6},
\ee
which also satisfies
\be
\bar\G^2 = 1,\qquad
\tr\bar\G = 0.
\ee
Notice that the first line of the dual 1+5 action \eqref{dual15action} may be obtained 
from the first line of the gauge-fixed PST action \eqref{PSTM5noncov} 
by the replacement rule 
\be
i(\Hbt\cdot u)^{\m\n}\to (H\cdot v)^{\m\n},\qquad 
i(H\cdot u)^{\m\n}\to (\Hbt\cdot v)^{\m\n}.
\ee 
This formal relation above between actions is a typical characterisation of 
a formulation and its dual. For example, the standard D$p$-brane actions in terms of 
their worldvolume vector fields are related to their electromagnetic 
dual counterparts, 
which is written in terms of the $(p-2)-$forms by a formal replacement rule similar to the above \cite{Pasti:1997gx, Aganagic:1997zk}. 
However, being a self-dual gauge theory, the M5-brane action \eqref{PSTM5noncov} 
is invariant under the worldvolume dualisation of the 2-form gauge field. 
Nevertheless, it is found in \cite{Maznytsia:1998yr, Maznytsia:1998xw} that the dualisation of linearised \eqref{PSTM5} with respect to the 
auxiliary field $a(x)$ gives the covariant form of the linearised \eqref{dual15action}. 
This is why we call \eqref{dual15action} the dual formulation of the 
M5-brane action. 
\cite{Maznytsia:1998yr, Maznytsia:1998xw} suggest that \eqref{dual15action} 
may be covariantized by an auxiliary 4-form. 
However, the covariantisation issue is quite complicated and we 
will not touch it upon throughout this paper.

The derivation and discussions on the M5-brane action in the dual formulation \eq{dual15action}
are presented in the subsequent sections.

\setcounter{equation}0 
\section{Derivation}   \label{sec:deri}
\subsection{EOM from superembedding}
The complete set of equations of motion of the action \eq{PSTM5}
has been shown \cite{Howe:1997vn, Bandos:1997gm}
to be equivalent to those obtained 
from the superembedding approach
\cite{Howe:1996yn}. In particular, when constructing a single M5-brane action
in the 3+3 formulation \cite{Ko:2013dka} and in the yet-incomplete 2+4 formulation \cite{Ko:2015zsy},
the chiral 2-form equations of motion obtained from the superembedding approach
provide useful information on how the action which gives the required equations of motion
should look like. 
As in the case of the other formulations, also in the dual 1+5 formulation, it is useful to discuss the equations of motion
of the chiral 2-form obtained from the superembedding approach.

In the superembedding  
formulation of the M5--brane \cite{Howe:1996yn,Howe:1997fb} the field strength $H_3$ of the chiral field $B_2$ is expressed in terms of
an auxiliary self--dual tensor $h_3=*h_3$ as follows
\footnote{Our normalisation convention of the field strength differs from that in \cite{Howe:1997vn} by the factor of $\frac 14$ in front of $H_3$.}
\be\label{embsd}
\frac 14 H_{\m\n\rho}=m^{-1\lambda}_{\mu}h_{\lambda\n\rho}\,, \qquad \frac 14\tilde H^{\m_1\n_1\rho_1}=\frac{1}6\epsilon^{\m_1\n_1\rho_1\m\n\rho}m^{-1\lambda}_{\mu}h_{\lambda\n\rho}=Q^{-1}m^{\mu_1\lambda}
h_{\lambda}{}^{\n_1\rho_1}\,
\ee
where $m^{-1\lambda}_{\mu}$ is the inverse matrix of
\be\label{mk}
m_{\mu}{}^{\lambda}=\delta_{\mu}{}^{\lambda}-2k_{\mu}{}^{\lambda}\,,\qquad m_\mu^{-1\lambda}=Q^{-1}(2\delta_{\mu}{}^{\lambda}-m_{\mu}{}^{\lambda}),\qquad k_{\mu}{}^{\lambda}=h_{\mu\nu\rho}h^{\lambda\nu\rho}\,
\ee
and
\be\label{Q}
Q=1-\frac 23 \tr \,k^2\,, \qquad 
\Ht^{\m\n\r} = \ove{3!}\e^{\m\n\r\a\b\g}H_{\a\b\g}. 
\ee
As was shown in \cite{Howe:1997vn}, by splitting the indices in eqs. \eqref{embsd} into 1+5 and expressing components of $h_3$ in terms of $H_{\mu\nu 5}$, one gets the duality relation (in our convention) 
\be\label{HtHsem}
\Ht = \frac{\lrbrk{1-\hlf\tr(H^2)}H + H^3}{\sqrt{1-\hlf\tr(H^2) + \ove{8}\tr(H^2)^2 - \ove{4}\tr(H^4)}}
\ee
where
$H$ and $\Ht$ are matrices with components
\be
H^\una{}_\unb\equiv H_5{}^\una{}_\unb,\qquad
\Ht^\una{}_\unb\equiv \Ht_5{}^\una{}_\unb.
\ee
Inverting the equation \eq{HtHsem} gives
\be\label{HHtsem}
H = \frac{\lrbrk{1+\hlf\tr(\Ht^2)}\Ht - \Ht^3}{\sqrt{1+\hlf\tr(\Ht^2) + \ove{8}\tr(\Ht^2)^2 - \ove{4}\tr(\Ht^4)}}.
\ee

Although both the equations \eq{HtHsem} and \eq{HHtsem}
are essentially the same, only the latter one arises directly,
as a consequence of Euler-Lagrange equation
for the $B_2$ sector (with all background fields and other worldvolume fields turned off)
of the action \eq{PSTM5noncov}, which first presented by 
\cite{Perry:1996mk, Pasti:1997gx}. 
On the other hand, to the best of our knowledge,
the Lagrangian which directly gives rise to the equation \eq{HtHsem}
has not appeared before in the literature, let alone its extended version
to describe the complete single M5-brane theory.
Thus in this paper, we construct and present
the complete single M5-brane theory in the form which serves this purpose.
This action is given in the equation \eq{dual15action}.

The construction of this action starts from
constructing
the flat space free theory and then
its nonlinearisation. 
To achieve the latter, we appeal to 
the Hamiltonian analysis and apply the idea of \cite{DeCastro:2002vd, DeCastro:2001he,DeCastro:1998pm}.

\subsection{Free theory in non-covariant form}   \label{sec:free}

Let us start by deriving the linearised version of \eqref{HtHsem} from an action principle. 

We would like to derive the linear self-duality condition
\be \label{LSD}
H^{\m\n\r} = \ove{3!}\e^{\m\n\r\t\s\l} H_{\t\s\l} = \Ht^{\m\n\r}
\ee
on the 3-form field strength $H_3=dB_2$ of a 2-form potential $B_2$ from a 6d Lagrangian. 
Consider the following 1+5 splitting of field strength, 
\be
H_{\m\n\r} = (H_{\underline{lmn}} , H_{\unm\unn5}), \qquad \unl,\unm,\unn=0,1,2,3,4. 
\ee
The Levi-Civita symbol is split according to 
\be
\e^{012345} = 1 = -\e_{012345}, \quad \Rightarrow\quad 
\e^{012345} = \e^{01234}\e^5 = \e^{01234} = -\e_{01234}\e_5=-\e_{01234}. 
\ee
The Greek letters are 6d indices $\m,\n,\cdots = 0,1,2,3,4,5$ while 
the underlined latin indices are 5d ones $\unl,\unm,\unn = 0,1,2,3,4$. 
Therefore, 
\be
\Ht^{\underline{lmn}}
= \ove{2!}\e^{\underline{lmnpq}}H_{\underline{pq}5}, \quad 
\Ht^{\underline{pq}5} 
= -\ove{3!}\e^{\underline{pq}\underline{lmn}}H_{\underline{lmn}}. 
\ee
The self-duality equation \eqref{LSD} could be derived from the following action:  
\be \label{ffdualS}
S = -\ove{4}\int d^6x 
\lrbrk{ H_{\underline{mn}5} \lrbrk{ H^{\underline{mn}}{}_{5} - \Ht^{\underline{mn}}{}_5 } }. 
\ee
The action has the following semi-local gauge symmetry 
\be  \label{SLGS}
\d B_{\unm\unn} = \Omega_{\unm\unn}(x^{\underline{k}}), \qquad 
\d B_{\unm5} = 0, 
\ee
where $\Omega_{\unm\unn}(x^{\underline{k}})$ are arbitrary functions of 5d coordinates $x^{\underline{k}}$. 
To be eligible as a gauge symmetry, the Noether charge associated with the semi-local symmetry must vanish at least on-shell. 
The conserved Noether current associated with \eqref{SLGS} is 
\be
j^\m = \ove{2}\lrbrk{H^{\underline{mn}5} - \Ht^{\underline{mn}5}} \Omega_{\unm\unn} \d^\m_5. 
\ee
It is clear that the Noether charge $Q=\int j^0 d^5x$ vanishes identically \textit{off-shell}, 
as $j^0=0$.  
Had we aligned the temporal direction in the `1' of `1+5' splitting, 
this would not be the case. In other words, the special direction chosen in the dual 1+5
formulation must be a spatial one. 

The equations of motion derived by varying \eqref{ffdualS} are 
\be  \label{ffEOM}
\begin{split} 
\pa_5\lrbrk{ \ove{3!}\e^{\underline{pqlmn}} (H_{\underline{lmn}} - \Ht_{\underline{lmn}}) } 
&= 0,  \\
\pa_{\underline{p}}\lrbrk{\ove{3!}\e^{\underline{pqlmn}} (H_{\underline{lmn}} - \Ht_{\underline{lmn}})} 
&= 0. 
\end{split}
\ee
The general solution to the field equations \eqref{ffEOM} is 
\be  \label{ffgnesoln}
\ove{3!}\e^{\underline{pqlmn}} (H_{\underline{lmn}} - \Ht_{\underline{lmn}}) 
= \ove{2!}\e^{\underline{pqlmn}} \pa_{\underline l} \o_{\unm\unn}(x^{\underline{k}}), 
\ee
where $\omega_{\underline{mn}}(x^{\underline k})$ are arbitrary functions of 5d coordinates 
$x^{\underline k}$. 

Notice that the components of the field strength $H_{\unm\unn5}$ are invariant under the transformation \eqref{SLGS}. Under \eqref{SLGS}, the left hand side of \eqref{ffgnesoln} transforms as 
\be
\d\lrbrk{ \ove{3!}\e^{\underline{pqlmn}} (H_{\underline{lmn}} - \Ht_{\underline{lmn}}) }
= \ove{2!}\e^{\underline{pqlmn}} \pa_{\underline{l}} \Omega_{\unm\unn}, 
\ee
which is in exactly the same form as the right hand side of \eqref{ffgnesoln}. 
Therefore, by the gauge-fixing $\Omega_{\unm\unn} = \omega_{\unm\unn}$, 
one obtains the self-duality equations 
\be
\ove{3!}\e^{\underline{pqlmn}} (H_{\underline{lmn}} - \Ht_{\underline{lmn}}) = 0 
\ee
which is obviously equivalent to \eqref{LSD}. 

The action \eqref{ffdualS} is manifestly invariant under the $SO(1,4)$ subgroup 
of the 6d Lorentz symmetry. 
However, although less obvious, it is also invariant under the following modified 
Lorentz transformation parametrized by $\L_{\underline{m}5} \equiv \L_{\underline{m}}$ mixing the $x^5$ and other directions $x^{\underline{m}}$: 
\be
\begin{split}
\d B_{\underline{mn}} 
&= \lrsbrk{(\L\cdot x)\del_5 - x^5(\L\cdot \del) } B_{\underline{mn}}
	- 2\L_{[\underline{m}}B_{\underline{n}]5}  
	+x_5\L^{\underline{l}} \lrbrk{H_{\underline{mnl}} - \Ht_{\underline{mnl}}}  	\\
\d B_{\underline{m}5}
&= \lrsbrk{(\L\cdot x)\del_5 - x^5(\L\cdot \del)} B_{\underline{m}5}
	-\L^{\underline{n}} B_{\underline{mn}},              \label{app:dBml}
\end{split}
\ee
where $(\L\cdot x) = \L_{\underline{m}5}x^{\underline{m}}$ 
and $(\L\cdot\pa) = \L^{\underline{m}}{}_5\pa_{\underline{m}}$. 
Therefore, the action \eqref{ffdualS} enjoys the full 6d Lorentz symmetry. 
The modified Lorentz symmetry reduces to the standard one when the 
field strength satisfies the self-duality equation. 

The free theory introduced
above
could be put on a curved 6d space. 
Consider the following action 
\be \label{cfdualS}
S = -\ove{4}\int \textrm{d}^6 x
	\lrbrk{ \frac{\sqrt{-g}}{g_{55}} 
		H_{5\underline{jk}} \lrbrk{H_5{}^{\underline{jk}} - \Hbt_5{}^{\underline{jk}} }
	}, 
\ee
where 
\be   \label{curvedHbt}
\Hbt^{\m\n\r} = \ove{3!\sqrt{-g}}\e^{\m\n\r\t\s\l} H_{\t\s\l}. 
\ee
Indices in \eqref{cfdualS} are pulled up and down by the 6d metric $g_{\m\n}$ 
with the mostly positive signature $(-+++++)$. 
The action is still invariant under the semi-local gauge symmetry \eqref{SLGS}. 
By varying the action \eqref{cfdualS}, we obtain the field equations 
\be   \label{cfEOM}
\begin{split}
\pa_5\lrbrk{ \e^{\underline{lmnpq}} \frac{4}{3g_{55}}g_{5[5}
		   (H - \Hbt)_{\underline{lmn}]} }
= 0, 	\\ 
\pa_{\underline{q}} \lrbrk{ \e^{\underline{lmnpq}} \frac{4}{3g_{55}}g_{5[5}
					(H - \Hbt)_{\underline{lmn}]} }
= 0. 
\end{split}
\ee
The general solution to \eqref{cfEOM} is 
\be \label{cfgensoln}
\e^{\underline{lmnpq}} \frac{4}{3g_{55}}g_{5[5}(H-\Hbt)_{\underline{lmn}]}
= \e^{\underline{pqijk}} \pa_{\underline{k}} \omega_{\underline{ij}}(x^{\underline{l}}), 
\ee
where $\omega_{\underline{ij}} = \omega_{\underline{[ij]}}(x^{\underline{l}})$ are arbitrary 
functions of 5d coordinates $x^{\underline{l}}$. 

One could obtain the self-duality equations 
\be
\e^{\underline{lmnpq}} \frac{4}{3g_{55}}g_{5[5}(H - \Hbt)_{\underline{lmn}]}
= 0 
\ee
by an appropriate gauge-fixing of the semi-local gauge symmetry \eqref{SLGS}. 

The action \eqref{cfdualS} enjoys the full 6d diffeomorphism. 
However, the diffeomorphism transformations of $\d_\e B_{\unm\unn}$ are modified 
in the directions $\xi^{\underline{l}}$. 
Indeed, after a somewhat lengthy algebra, one
shows that the action \eqref{cfdualS} 
is invariant (up to total derivative terms) under 
\be
\begin{split}
\d_\e B_{\underline{ij}}
&=  \xi^\m H_{\m \underline{ij}} 
	- 4\frac{\xi^{\underline{p}}}{g_{55}}g_{5[5} 
		\lrbrk{H_{\underline{ijp}]} - \Hbt_{\underline{ijp}]}}   \\
&= \xi^\m H_{\m \underline{ij}}
	+ \frac{\sqrt{-g}}{2!}\e_{\underline{ijpmn}} 
	\frac{\xi^{\underline{p}}}{g_{55}} 
	\lrbrk{H^{\underline{mn}}{}_5 - \Hbt^{\underline{mn}}{}_5}, \\
\d_\e B_{5\underline{m}}
&= \xi^\m H_{\m5\underline{m}}. 
\end{split}
\ee
In the next subsection, we will generalise \eqref{cfdualS} to a nonlinear theory 
following the idea of \cite{DeCastro:2002vd, DeCastro:2001he,DeCastro:1998pm}.

\subsection{Dual 1+5 Lagrangian from Hamiltonian}  \label{sec:deC}

The  
HT1+5 nonlinear theory, 
which is the chiral 2-form part of PST M5 action \eqref{PSTM5} with the gauge-fixing $a = x^0$, contains primary constraints 
$\Ht^{0\ah\bh} + \p^{\ah\bh} = 0$ ($\ah,\bh=1,2,3,4,5$), 
where $\p^{\ah\bh}$ are conjugate momenta to $B_{\ah\bh}$. 
\cite{DeCastro:2002vd, DeCastro:2001he,DeCastro:1998pm} showed that one could obtain the  
PS1+5 nonlinear 
theory, 
which is the chiral 2-form part of \eqref{PSTM5noncov}, 
if one replaces $\Ht^{0\ah\bh}$ in the  
HT1+5 Hamiltonian by $(\Ht^{0\ah\bh}-\p^{\ah\bh})/2$ 
and then relax the primary constraints $\Ht^{0\ah\bh} + \p^{\ah\bh} = 0$ 
to $\Ht^{0i5} + \p^{i5} = 0$ ($i=1,2,3,4$). 
The manifest $SO(5)$ covariant form of  
HT1+5 formulation would be first decomposed 
to $SO(4)$ by relaxing the primary constraints, and then the indices $0$ and $i$ would be recombined to get a 
PS1+5 Lagrangian with the manifest $SO(1,4)$ covariance. 
In this section, we will apply the similar technique by relaxing the primary constraints 
$\Ht^{0\ah\bh} + \p^{\ah\bh} = 0$ to $\Ht^{0ij} + \p^{ij} = 0$. 
The result is expected to be the nonlinear dual 1+5 Lagrangian written 
in terms of the components $H_{5\underline{mn}}$, where $\unm,\unn=0,1,2,3,4$.

Let us start from HT1+5 Lagrangian 
\be
\cL = -\ove{4}\frac{\Ht^{0\ah\bh}H^0{}_{\ah\bh}}{g^{00}}
- 
\sqrt{-g}\sqrt{\det\lrbrk{\d_{\ah}^\bh + \frac{1}{\sqrt{\g}}\Ht^0{}_{\ah}{}^\bh}}, 
\ee
which is manifestly $SO(5)$ covariant and is already in a first-order form in the gauge field $B_2$. 
The $\Ht^0{}_{\ah}{}^{\bh}$ is defined as $\Ht^0{}_{\ah}{}^\bh \equiv \Ht^{0\m\bh}g_{\m\ah}$. 
For the future convenience of this section, let us rescale the Lagrangian
$\cL\to\cL' = 4\cL$ and then put it in the first-order form:
\be\label{HT15firstorder}
\begin{split}
\cL' &= \p^{\ah\bh}H_{0\ah\bh}+\Ht^{0\ah\bh}H_{\ch\ah\bh}N^\ch
- 4N\sqrt{\g}
\sqrt{
1-\frac{1}{2\g}\tr\Ht^2 - \frac{1}{4\g^2}\tr\Ht^4 + \frac{1}{8\g^2}(\tr\Ht^2)^2}  \\
&\quad +\xi_{\ah\bh}\lrbrk{\p^{\ah\bh}+\Ht^{0\ah\bh}}
\end{split}
\ee
where
\be
\tr\Ht^2\equiv\Ht^0{}_\ah{}^\bh\Ht^0{}_\bh{}^\ah,\qquad
\tr\Ht^4\equiv\Ht^0{}_\ah{}^\bh\Ht^0{}_\bh{}^\ch\Ht^0{}_\ch{}^\dh\Ht^0{}_\dh{}^\ah, 
\qquad 
\ee
and the metric is Arnowitt-Deser-Misner decomposed 
\be\label{metric}
g_{\m\n} =
\begin{pmatrix}
-N^2+\g_{\ah\bh}N^\ah N^\bh & \g_{\bh\ch}N^\ch\\
&\\
\g_{\ah\ch}N^\ch & \g_{\ah\bh}
\end{pmatrix}
.
\ee
We define the inverse of $\g_{\ah\bh}$ and its determinant as $\g^{\ah\bh},$ and $\g,$ respectively.
The inverse of the metric is
\be
g^{\m\n} =
\begin{pmatrix}
-N^{-2} & \frac{N^\bh}{N^2}\\
\frac{N^\ah}{N^2} & \g^{\ah\bh}-\frac{N^\ah N^\bh}{N^2}
\end{pmatrix}
.
\ee

In \cite{DeCastro:2002vd, DeCastro:2001he,DeCastro:1998pm}, the PS1+5 Hamiltonian can be obtained
from the HT1+5 Hamiltonian by first modifying
\be
\Ht^{0\ah\bh}\to \frac{\Ht^{0\ah\bh} - \p^{\ah\bh}}{2},
\ee
and then modifying the constraint $\p^{\ah\bh}+\Ht^{0\ah\bh} = 0$ to
\be
\p^{i5} = - \Ht^{0i5}.
\ee
Using the obtained form of the Hamiltonian,
then the PS1+5 Lagrangian can be obtained.

For us, we wish to obtain the dual 1+5 Lagrangian
by following the similar way.
So, we start from the
first-order Lagrangian given in the equation \eq{HT15firstorder}.
Let us modify
\be
\Ht^{0\ah\bh}\to \tilde\cF^{\ah\bh}\equiv\frac{\Ht^{0\ah\bh} - \p^{\ah\bh}}{2}.
\ee
Then modify the constraint $\p^{\ah\bh}+\Ht^{0\ah\bh} = 0$ to
\be
\p^{ij} = - \Ht^{0ij}.
\ee
This gives
\be
\tilde{\cF}^a{}_b = \frac{\g_{b5}}{\g_{55}}\tilde{\cF}^a{}_5 + \lrbrk{\g_{bc} - \frac{\g_{c5}\g_{b5}}{\g_{55}}}\Ht^{0ac}
\ee
\be
\tilde{\cF}^5{}_b = -\frac{\g_{bc}}{\g_{55}}\tilde{\cF}^c{}_5 + \frac{\g_{bc}}{\g_{55}}\Ht^{0cd}\g_{d5}
\ee
\be
\tilde{\cF}^5{}_5 = -\frac{\g_{c5}}{\g_{55}}\tilde{\cF}^c{}_5
\ee
\be
\tilde{\cF}^a{}_5 = \tilde{\cF}^a{}_5, 
\ee
where $\tilde{\cF}^\ah{}_\bh = \tilde{\cF}^{\ah\ch}\g_{\ch\bh}$ 
and we have intended to maintain the matrix form of $\tilde{\cF}$ which has 
one index up and one index down. 
Define $\pt^{\ah\bh} = \pt^{[\ah\bh]}$ such that
\be
\pt^{ab} = \pt^{55} = 0,\qquad
\pt^{5a} = -\pt^{a5} = \frac{\tilde{\cF}^a{}_5}{g_{55}}
\ee
and $\Tt_\bh{}^\ah$ as
\be
\Tt_b{}^a = \lrbrk{\g_{bc} - \frac{\g_{c5}\g_{b5}}{\g_{55}}}\Ht^{0ac},\qquad
\Tt_b{}^5 = \frac{\g_{bc}}{\g_{55}}\Ht^{0cd}\g_{d5},\qquad
\Tt_5{}^\ah = 0, 
\ee
so that we have 
\be
\tilde{\cF}^\ah{}_\bh = g_{\bh\ch} \pt^{\ch\ah} + \Tt_\bh{}^\ah
= \pt_\bh{}^\ah + \Tt_\bh{}^\ah. 
\ee
The matrix $\Tt_{\bh}{}^{\ah}$ encodes the components $H_{5ab}$, 
while $\pt^{5a}$ contains the conjugate momenta $\p^{a5}$ 
which must be replaced with its equation of motion in order to get the Lagrangian of the theory. 

It is natural to expect the resulting dual 1+5 theory is described by the 
$SO(1,4)$ covariant tensor $H_{5\underline{ab}}$ ($\una,\unb=0,1,2,3,4$). 
Therefore, following the similar spirit of decomposition, we do the 
$SO(1,4)\rar SO(4)$ decomposition by writing 
\be
H_5{}^a{}_b = \frac{g^{a0}}{g^{00}}H_5{}^0{}_b + \g^{ac}H_{5cb}
\ee
\be
H_5{}^a{}_0 = -\frac{g^{ab}}{g^{00}}H_5{}^0{}_b - \g^{ab}N^c H_{5cb}
\ee
\be
H_5{}^0{}_0 = -\frac{g^{0a}}{g^{00}}H_5{}^0{}_a
\ee
\be
H_5{}^0{}_b = H_5{}^0{}_b. 
\ee
Let us define $p_{\una\unb} = p_{[\una\unb]}$ such that
\be
p_{ab} = p_{00} = 0,\qquad
p_{0a} = -p_{a0} = \frac{H_5{}^0{}_a}{g^{00}}
\ee
and $T^{\una}{}_{\unb}$ such that
\be
T^0{}_\unb = 0,\qquad
T^a{}_b = \g^{ac}H_{5cb},\qquad
T^a{}_0 = -\g^{ab}N^c H_{5cb}.
\ee
Therefore,
\be
H_5{}^\una{}_\unb = g^{\una\unc}p_{\unc\unb} + T^{\una}{}_{\unb}.
\ee

After performing our version of \cite{DeCastro:2002vd, DeCastro:2001he,DeCastro:1998pm} procedure, 
the full nonlinear Lagrangian is then
\be\label{unfinishedfirstorder}
\begin{split}
\cL' &= -\Ht^{0ab}H_{0ab}
+2\Ht^{0a5}H_{0a5}
-4\pt^{a5}p_{a0}
+4\frac{\Ht^{0ac}\g_{c5}H^0{}_{a5}}{\g_{55}g^{00}}
+\ove{2}\Ht^{0ab}\e_{abmn}\Ht^{0mn}N^5\\
&\qquad-4N\sqrt{\g}\sqrt{\cA + \cC_{ab}\pt^{a5}\pt^{b5}},
\end{split}
\ee
where
\be
\cA =
1- \frac{1}{\g}\Tt_a{}^{[b} \Tt_b{}^{a]}
-\frac{1}{\g^2}\Tt_a{}^{[b} \Tt_b{}^{c}\Tt_c{}^{d}\Tt_d{}^{a]}
\ee
\be
\cC_{ab} =\lrbrk{-g_{55}g_{mb}+g_{m5}g_{b5}}
\lrbrk{
- \frac{1}{\g}\d_a^m
-\frac{1}{\g^2}\bigg(\Tt_{a}{}^n \Tt_n{}^m
-\hlf\d_a^m\Tt_i{}^j \Tt_j{}^i\bigg)
}.
\ee
To get the final form of the Lagrangian, we need to replace $\pt^{a5}$ (which contains the conjugate momenta $\p^{a5}$) with its equation of motion. 
The equation of motion of $\pt^{a5}$ is
\be
\begin{split}
0=\frac{\d\cL}{\d\pt^{a5}}
&=
-4p_{a0}-4\frac{N\sqrt{\g}}{\sqrt{\cA + \cC_{ab}\pt^{a5}\pt^{b5}}}(\cC_{ab}\pt^{b5})
,
\end{split}
\ee
or
\be  \label{eq:EOMpt}
\begin{split}
0
&=p_{a0}
+\frac{N\sqrt{\g}}{\sqrt{\cA + \cC_{ab}\pt^{a5}\pt^{b5}}}(\cC_{ab}\pt^{b5}). 
\end{split}
\ee

Using  
the above equation of motion
to rewrite $-4\pt^{a5}p_{a0}$
in the Lagrangian. This gives
\be
\begin{split}
\cL' &= -\Ht^{0ab}H_{0ab}
+2\Ht^{0a5}H_{0a5}
+\ove{2}\Ht^{0ab}\e_{abmn}\Ht^{0mn}N^5
- 4\ove{g^{00}}H^0{}_{d5}\Ht^{0nd}\frac{\g_{n5}}{\g_{55}}
\\
&\qquad-4\frac{\cA N\sqrt{\g}}{\sqrt{\cA + \cC_{ab}\pt^{a5}\pt^{b5}}}\\
&= \frac{\sqrt{-g}}{g_{55}}H_{5\una\unb}\Hbt_5{}^{\una\unb}
-4\frac{\cA N\sqrt{\g}}{\sqrt{\cA + \cC_{ab}\pt^{a5}\pt^{b5}}}
\end{split}
\ee

By noting
\be
T^a{}_m T^m{}_n \g^{nb}
=
\hlf\frac{\g_{55}}{\g}\Tt_m{}^n\Tt_n{}^m\g^{ab}
-\frac{\g_{55}}{\g}\g^{am}\Tt_m{}^n\Tt_n{}^b,
\ee
and using \eqref{eq:EOMpt}
we can obtain
\be
\cL' =
\frac{\sqrt{-g}}{g_{55}}H_{5\una\unb}\Hbt_5{}^{\una\unb} 
- 4N\sqrt{\g}\sqrt{X + Z^{ab}p_{a0}p_{b0}},
\ee
where
\be
X = 
1- \frac{1}{g_{55}}T^a{}_{[b} T^b{}_{a]}
- \frac{1}{g_{55}^2} T^a{}_{[b} T^b{}_{c}T^c{}_{d}T^d{}_{a]},
\ee
\be
Z^{ab} =
- \frac{1}{g_{55}}\frac{\g^{ab}}{N^2}
- \frac{1}{g_{55}^2}\lrbrk{\ove{N^2}T^a{}_m T^m{}_n\g^{nb} -\hlf\ove{N^2}\g^{ab}T^m{}_i T^i{}_m }
\ee

Note that 
the manifestly $SO(4)$ covariant expression inside the square root 
can be assembled back to a nice $SO(1,4)$ covariant form
\be
X + Z^{ab}p_{a0}p_{b0}
=
1
- \frac{1}{g_{55}} H_{5}{}^\una{}_{[\unb} H_{|5|}{}^{\unb}{}_{\una]}
- \frac{1}{g_{55}^2} H_5{}^\una{}_{[\unb} H_{|5|}{}^{\unb}{}_{\unc}H_{|5|}{}^{\unc}{}_{\und}H_{|5|}{}^{\und}{}_{\una]}. 
\ee
The nonlinear dual 1+5 action 
is then 
\be  \label{cfdualSnlraw}
S' =
\int \textrm{d}^6 x\ \lrbrk{ \frac{\sqrt{-g}}{g_{55}}H_{5\una\unb}\Hbt_5{}^{\una\unb}
- 4\sqrt{-g}\sqrt{\det\lrbrk{\d_{\una}^{\unb} + \frac{1}{\sqrt{g_{55}}}H_{5\una}{}^{\unb}}}
},
\ee
or
\be
\label{cfdualSnl}
S =
\int \textrm{d}^6 x\ \lrbrk{
\ove{4}\frac{\sqrt{-g}}{g_{55}}H_{5\una\unb}\Hbt_5{}^{\una\unb}
- \sqrt{-g}\sqrt{\det\lrbrk{\d_{\una}^{\unb} + \frac{1}{\sqrt{g_{55}}}H_{5\una}{}^{\unb}}}
}.  
\ee
As the nonlinear part of the action depends on the field strength through $H_{5\una\unb}$, the action still enjoys the semi-local gauge symmetry \eqref{SLGS}. 
As a result, the procedures of gauge-fixing to get the self-duality equations follows 
exactly the steps presented for the free theory. 
The upshot is that field equations of the above action is equivalent to the 
nonlinear self-duality equations \eqref{nlSD}. 

Note that
the action \eqref{cfdualSnl}
enjoys the full 6d diffeomorphism invariance. 
However, the diffeomorphism transformations of $\d_\e B_{\underline{m}\underline{n}}$ are modified 
in the directions $\xi^{\underline{l}}$. 
Indeed, after a somewhat lengthy algebra, one 
shows that the action
\eqref{cfdualSnl} 
is invariant (up to total derivative terms) under 
\be  \label{moddiffnl}
\begin{split}
\d_\e B_{\underline{ij}}
&= \xi^\m H_{\m\underline{ij}}
	- \xi^{\underline{p}}\lrbrk{ 4\ove{g_{55}}g_{5[5}H_{\underline{ijp}]} + 3
	\e_{\underline{abijp}} \frac{\d V}{\d H_{\una\unb5}} }, \\
\d_\e B_{5\underline{m}}
&= \xi^\m H_{\m5\underline{m}}. 
\end{split}
\ee
The above transformations reduce to the standard diffeomorphism rules 
if the self-duality equations are satisfied. 

In this subsection, we have obtained the dual 1+5 Lagrangian from the Hamiltonian.
However, in order for this derivation to be justified, one of the requirements
is to show that the theory has the correct number of degrees of freedom.
In the next section, we will show that this is indeed the case.

\setcounter{equation}0
\section{Constraint analysis}   \label{sec:constr}
Recall that from
section \ref{sec:deC}, we have used the first-order Lagrangian for dual 1+5 theory
\be\label{curvednld15}
\begin{split}
\cL'
&= \p^{\ah\bh}H_{0\ah\bh}+\hlf\tilde\cF^{\ah\bh}\tilde\cF^{\xh\yh}\e_{\ah\bh\ch\xh\yh}N^\ch  \\
&\quad - 4N\sqrt{\g}
\sqrt{
1-\frac{1}{2\g}\tr\tilde\cF^2 - \frac{1}{4\g^2}\tr\tilde\cF^4 + \frac{1}{8\g^2}(\tr\tilde\cF^2)^2
}  \\
&\quad +\xi_{ij}(\p^{ij}+\Ht^{0ij})
\end{split}
\ee
to arrive at the non-linear dual 1+5 action \eq{cfdualSnlraw}.
By reading off from the first-order Lagrangian \eq{curvednld15},
the Hamiltonian and momentum densities are then given by
The Hamiltonian and momentum densities are then given by
\be
\begin{split}
\cH_0
&=
4\sqrt{\g}\sqrt{1-\ove{2\g}\tr \tilde\cF^2 - \ove{4\g^2}\tr \tilde\cF^4 + \ove{8\g^2}(\tr \tilde\cF^2)^2},
\end{split}
\ee
\be
\cH_\ch = -\hlf\tilde\cF^{\ah\bh}\tilde\cF^{\xh\yh}\e_{\ah\bh\ch\xh\yh}.
\ee

After combining the Lagrangian \eq{curvednld15} with that of 6d gravity,
one then follows for example the Dirac constraint analysis \cite{Dirac:1950pj, Henneaux:1992ig}. 
Finally,
One obtains the 
first-class constraints:
\be
\P_\m\approx 0\ (6),\qquad
\p^{0\ah}\approx 0\ (5),\qquad
\cH_\m\approx 0\ (6),\qquad
\pa_\ah\p^{\ah\bh}\approx 0\ (4), 
\ee
where $\P_\m$ is the conjugate momenta of $N^\m$, and the 
second-class constraints:
\be
\Ht^{0ab}+\p^{ab}\approx 0\ (6).
\ee
In total, the nonlinear dual 1+5 chiral 2-form theory has 
$21$ first-class constraints and $6$ second-class constraints. 
As there are $72$ phase space variables (30 from $B_{\m\n}$ and $\p^{\m\n}$ 
and $42$ from $g_{\m\n}$ and $\P^\m,\zeta^{\mh\nh}$), 
the
number of degrees of freedom is then
\be
\frac{72-2\times 21 - 6}{2} = 12 = 9+3.
\ee
We thus have 9 propagating degrees of freedom for graviton 
and 3 for chiral 2-form. 

By using the identities
\be
(\tilde\cF^3)_{\ah\bh}\e^{\ph\ah\bh\xh\yh}\tilde\cF_{\xh\yh}
=\ove{4}\tr(\tilde\cF^2)\e^{\ph\xh\yh\mh\nh}\tilde\cF_{\mh\nh}\tilde\cF_{\xh\yh},
\ee
\be
(\tilde\cF^3)_{\ah\bh}\e^{\ph\ah\bh\xh\yh}(\tilde\cF^3)_{\xh\yh}
=-\lrbrk{\ove{4}\tr(\tilde\cF^4)-\ove{8}\tr(\tilde\cF^2)^2}
\e^{\ph\mh\nh\xh\yh}\tilde\cF_{\mh\nh}\tilde\cF_{\xh\yh},
\ee
it can be shown that
the hypersurface deformation algebra 
\cite{Hojman:1976vp}
\bea 
[\cH^{\text{full}}_0(x),\cH^{\text{full}}_{0}(x')] &=& (\g^{\ah\bh}(x)\cH^{\text{full}}_\ah(x)+\g^{\ah\bh}(x')\cH^{\text{full}}_\ah(x'))\pa_\bh\d^{(5)}(x,x'),\\\
[\cH^{\text{full}}_\ah(x),\cH^{\text{full}}_0(x')] &=& \cH^{\text{full}}_0(x)\pa_\ah\d^{(5)}(x,x') +\pa_{\mh}\p^{\mh\nh}(x)\frac{\d\cH_0}{\d\tilde\cF^{\nh\ah}}(x)\d^{(5)}(x,x'),\\
\
[\cH_\ah^{\text{full}}(x),\cH^{\text{full}}_\bh(x')] &=& \cH^{\text{full}}_\ah(x')\pa_\bh\d^{(5)}(x,x')+\cH^{\text{full}}_\bh(x)\pa_\ah\d^{(5)}(x,x')\nn \\
&&\quad +\pa_{\mh}\p^{\mh\nh}(x)\e_{\bh\jh\kh\ah\nh}(x)\tilde\cF^{\jh\kh}(x)\d^{(5)}(x,x'),
\eea
where $\cH^{\text{full}}_\m = \cH^{(g)}_\m + \cH_\m,$
is satisfied.
The pure gravity energy and momentum densities are given by 
\be
\cH^{(g)}_0 = -\sqrt{\g} R + \ove{\sqrt{\g}} \lrbrk{ \zeta^{\ih\jh}\zeta^{\mh\nh}\g_{\ih\mh}\g_{\jh\nh} - \ove{2}(\zeta^{\mh\nh}\g_{\mh\nh})^2 }, \qquad 
\cH^{(g)}_{\mh} = -2\g_{\mh\nh} \nabla_{\ph} \zeta^{\nh\ph}, 
\ee
where $\g_{\mh\nh}$ is spatial 5d metric, $\zeta^{\mh\nh}$ is the conjugate momenta 
to $\g_{\mh\nh}$, $R$ is 5d Ricci scalar and $\nabla_{\mh}$ is $\g$-compatible 
covariant derivative. 

By following for example the procedures outlined in \cite{Henneaux:1992ig},
it can be shown that the Hamiltonian density $\cH^{\text{full}}_0$ and momentum densities $\cH^{\text{full}}_\ah$
generate the modified diffeomorphism transformation presented in equation
\eq{moddiffnl}.

\setcounter{equation}0
\section{Comparison of on-shell actions}   \label{sec:osS}
Although the duality-symmetric actions corresponding to formulations with different splittings of space-time are different off-shell, they should agree with each other on-shell. For free chiral 2-form theories, it was found that the free theory actions with different splittings all vanish on-shell in \cite{Pasti:2009xc}. 
In \cite{Ko:2013dka}, it was shown that the chiral 2-form part of both 1+5 and 3+3 M5-brane action agree with each other and is given by 
\be  \label{onshellS}
S^{(\text{on-shell})}
= -\int d^6x \sqrt{-g} \frac{2}{Q} 
 + \ove{2}\int_{\mathcal{M}_6} \lrbrk{ C_6 + H_3 \wedge C_3 } 
\ee
on-shell\footnote{The action in \cite{Ko:2013dka} is twice as our \eqref{PSTM5} in the convention of overall numerical normalisation.}. 
Physically, the on-shell value of the chiral 2-form part of the M5 action determines 
the tension of the string soliton \cite{Perry:1996mk}. 
This on-shell property of duality-symmetric action was used in \cite{Ko:2015zsy} to obtain the nonlinearisation of the 2+4 action. 

To put the dual 1+5 M5-brane action on-shell, the superembedding equations 
\be
\begin{split}
\Ht_{\underline{ab}5}
= 4Q^{-1}\lrbrk{(1-2\tr f^2) f_{\una\unb} + 8(f^3)_{\una\unb}}, \\
H_{\underline{ab}5}
= 4Q^{-1}\lrbrk{(1+2\tr f^2)f_{\una\unb} - 8(f^3)_{\una\unb}}, 
\end{split}
\ee
where $f_{\una\unb} = h_{\una\unb5}$, are substituted into the M5-brane action. 
We found that the on-shell dual 1+5 M5-brane action is also given by \eqref{onshellS},
and hence it agrees with the on-shell actions for the 1+5 and 3+3 cases,
despite the fact that all of them have different off-shell actions
from one another.

The off-shell differences are of interests because their understanding may shed 
some light on the issue of quantising self-dual fields. However, this is still an open problem.

\setcounter{equation}0
\section{Double dimensional reduction}   \label{sec:DDR}
It is known that M--theory on a circle is dual to type IIA string theory \cite{Townsend:1995kk, Witten:1995ex}. 
Indeed, if one wraps the M5--brane on the compact direction, 
one expects D4--brane be obtained. 
This is called double dimensional reduction because both dimensions of the worldvolume of the M5--brane as well as the target space are reduced. 
It was shown in \cite{Pasti:1997gx, Aganagic:1997zq} that the gauge-fixed PST 
M5 action gives rise to the dual D4--brane action \cite{Aganagic:1997zk} upon 
double dimensional reduction. 
In this section, we will show that the dual 1+5 M5 action reduces 
to the standard D4--brane action written in terms of the worldvolume vector gauge field directly.

Let $X^{10}$ be the compact direction that $x^5$ wraps on. 
After the dimensional reduction, only the zero Fourier modes are kept. 
For simplicity, let us consistently neglect the vector and scalar fields that 
arise from the reduction of the metric tensor. 
In particular, the various objects reduce according to 
\be
\begin{split}
H_{\m\n\r} &\rar (H_{\underline{mnp}} , F_{\unm\unn}), \\
g_{\m\n} &\rar g_{\unm\unn}, \\
C_{\m\n\r} &\rar (C_{\underline{mnp}} , C_{\unm\unn}), \\
C_{\m_1\cdots\m_6} &\rar  C_{\unm_1\cdots\unm_5}, \\
v_\l &\rar \d_\l^5, \qquad v^\l \rar \d^\l_5, 
\end{split}
\ee
where we have defined 
\be
H_{\unm\unn5} \equiv F_{\unm\unn}, \quad 
C_{\unm\unn5} \equiv C_{\unm\unn}, \quad 
C_{\unm_1\cdots\unm_55} \equiv C_{\unm_1\cdots\unm_5}. 
\ee
The $C_{\underline{mnp}}$ 
is the Ramond-Ramond 3-form 
while $C_{\underline{mn}}$  
serves as Kalb-Ramond field in string theory. 

A straightforward computation leads to 
\be  \label{D4S}
\begin{split}
S_5
&= -\int d^5x \sqrt{-\det(g_{\unm\unn} + F_{\unm\unn})} 
	- \int_{\cM_5} \lrbrk{ e^{F_2} \wedge (C_3 + C_5')  
					}_5, 
\end{split}
\ee
where 
\be
C_5' = \ove{2}C_5 - \ove{2}C_2\wedge C_3, 
\ee
\be
F_{\unm\unn} = \pa_{\unm}A_{\unn} - \pa_{\unn}A_{\unm} + C_{\underline{mn}}
\ee
is the extended field strength with $A_{\una}\equiv B_{\una5}$. 
The $H_{\underline{mnp}}$ components appear in total derivative terms 
after reduction and hence are discarded. 
As a result, only $B_{\una5}$ components of the chiral 2-form survives 
and serves as the vector gauge field $A_{\una}$ in the D4 worldvolume.  
The Wess-Zumino term is written in a formal manner that only 
5-forms out of the wedge product of $(C_3+C_5')$ with the formal expansion of $\exp(F_2)$ are integrated. 

The D4--brane action \eqref{D4S} obtained by double dimensional reduction of dual 1+5 M5 action is in a standard form \cite{Douglas:1995bn, Green:1996bh}. 
It is obtained by trivial computation without the need to further dualise 
any resulting worldvolume gauge field.

\setcounter{equation}0
\section{Conclusion}\label{sec:conc}

	We have constructed a dual 1+5 formulation with respect to the conventional PST formalism for the single M5--brane action propagating in a generic 11d supergravity background. The dual 1+5 M5 action has both the required local gauge symmetries on the worldvolume, i.e. general coordinate diffeomorphism invariance and kappa symmetry, although the action is in a non-manifestly covariant form. 
To equalise the field equations and the self-duality conditions, 
a semi-local gauge symmetry is utilised. 
In order for this semi-local symmetry to be eligible as a gauge symmetry, 
the special direction singled out from 6d must be spatial. 
Similar restrictions on the choices of temporal direction from the subspaces of 
splittings of worldvolume space was also observed in 2+4 formulation \cite{Ko:2015zsy}. The use of semi-local gauge symmetry is also necessary 
for theories of chiral forms in topologically nontrivial space-time \cite{Bekaert:1998yp, Bandos:2014bva, Isono:2014bsa}. 
The dual 1+5 M5--brane formulation will be even more useful if we could validate its usage on topologically nontrivial worldvolume in the future.

The construction of the dual 1+5 M5 action starts from the free theory. 
The detailed analysis and gauge-fixing in Lagrangian formalism is presented. 
To nonlinearise the free theory, we followed the idea outlined in \cite{DeCastro:2002vd, DeCastro:2001he,DeCastro:1998pm} by switching to Hamiltonian of HT1+5 
and then relaxing certain constraints and finally completing the Legendre transformation. After obtaining the nonlinear dual 1+5 theory in curved 6d space in one go, we coupled the theory to background supergravity and found the kappa symmetry completion. 
Extending the idea of \cite{DeCastro:2002vd, DeCastro:2001he,DeCastro:1998pm}, 
it will be remarkable if one could find other possible formulations of self-interacting chiral 2-forms in a curved space by relaxing 
the primary constraints of HT1+5 in a different manner.

We computed the on-shell value of the dual 1+5 M5 action and found that 
it is written in terms of the superembedding scalar variable as in the case 
of its counterpart formulations, albeit the M5 actions in different splittings disagree 
with each other off-shell. 
On the other hand, we performed the double dimensional reduction on a circle 
and showed that dual 1+5 M5 action reduces directly to the standard conventional D4--brane action. This is in contrast to the conventional PST M5 action 
for which the double dimensional reduction results in a dual D4 action.

	The dual 1+5 action presented here is in a non-manifestly covariant form. 
	Although the splitting of 6d worldvolume by picking up a special direction 
	is similar to the case of conventional 1+5 formalism, 
	the PST covariantisation procedure \cite{Pasti:1995tn, Pasti:1995ii, Pasti:1996va, Pasti:1996vs}
	with a scalar field seems to be not doable. 
	Instead, an auxiliary 4-form is suggested by \cite{Maznytsia:1998yr, Maznytsia:1998xw} to covariantise the theory. 
	However, we found this issue to be more nontrivial than we thought 
	at this stage, and we will leave it as a possible future work. 
	Similar obstacles in the PST covariantisation was also found in the 2+4 
	formulation \cite{Ko:2015zsy}. 
	We hope to report progress on these issues in the near future. 
	
	In \cite{Schwarz:2013wra}, the PST M5--brane action in the background 
	of $AdS_7\times S^4$ is regarded as the \textit{exact} effective action 
	(called highly effective action there) of the (2,0) superconformal field theory 
	in the Coulomb branch. 
It will be interesting to verify whether the dual 1+5 M5 action 
	as well as the 3+3 M5 action \cite{Ko:2013dka} satisfy all the requirements \cite{Schwarz:2013wra} to be a highly effective action. 
	Moreover, it will be interesting as well to see how these off-shell different actions 
	capture the same quantum nature of the (2,0) superconformal field theory.

\subsection*{Acknowledgements}
We are very grateful to Dmitri Sorokin for 
various helpful  
remarks and comments on the manuscript.

The authors are also grateful to Chong-Sun Chu and Douglas Smith for discussions. 

Sh-L. K. would like to acknowledge the hospitality of department of electro-physics of National Chiao-Tung University 
during his visit at the final stage of this paper.

P.V. would like to acknowledge hospitality and support extended to him by
National Center for Theoretical Sciences, Physics Division, Taiwan
during the final stage of this paper.

\appendix

\setcounter{equation}0

\providecommand{\href}[2]{#2}\begingroup\raggedright\endgroup

\end{document}